\documentclass[fleqn,usenatbib]{mnras}

\usepackage{newtxtext,newtxmath}
\usepackage[T1]{fontenc}

\DeclareRobustCommand{\VAN}[3]{#2}
\let\VANthebibliography\thebibliography
\def\thebibliography{\DeclareRobustCommand{\VAN}[3]{##3}\VANthebibliography}

\usepackage{graphicx}	% Including figure files
\usepackage{amsmath}	% Advanced maths commands
\usepackage{lipsum}
\usepackage{url}
\usepackage[normalem]{ulem}
\usepackage{longtable}
\usepackage{booktabs}
\usepackage{pdflscape} % Landscape pages
\usepackage{array}
\def\aap{A \& A}

\def\apjs{ApJS}
\def\apjl{ApJL}

\def\la{\left\langle}

\title[Faraday Depolarization in LoTSS-DR2 galaxies]{Faraday Complexity and Depolarization in LOFAR Two-metre Sky Survey (LoTSS‑DR2) Polarized Radio Sources}

\author[R. Sinha et al.]{
Rudra Sinha,$^{1}$\thanks{E-mail: sinharudrairon@gmail.com}
and Abhik Ghosh$^{1}$\thanks{E-mail: abhik.physicist@gmail.com}
\\
$^{1}$Department of Physics, Banwarilal Bhalotia College, Asansol, West Bengal, Pin: 713303, India\\
}

\pubyear{\the\year{}}

\begin{document}
\label{firstpage}
\pagerange{\pageref{firstpage}--\pageref{lastpage}}
\maketitle

\begin{abstract}
 We present a broadband spectro‑polarimetric analysis of 1,565 polarized radio sources from the LOFAR Two‑Metre Sky Survey Data Release 2 (LoTSS‑DR2) RM Grid catalogue. This study uses frequency‑dependent Stokes $Q$ and $U$ spectra across the 120–168 MHz LOFAR HBA band to investigate their polarization properties. The polarization behaviour of each source is modelled with multi‑component Faraday depolarization models to investigate the magneto‑ionic environments responsible for low‑frequency depolarization. Significant Faraday complexity is observed throughout the sample, with 43.2\% of sources requiring two or three Faraday components. External Faraday dispersion dominates the depolarization behaviour, with 54.1\% of sources classified as external‑screen dominated and approximately 60\% showing statistically significant evidence for turbulent external Faraday‑active media, while only 10.3\% are consistent with pure internal differential Faraday rotation. The intrinsic polarization angle and RM separations between fitted components are generally small, suggesting that many RM components trace physically related emission regions embedded within common magneto‑ionic environments. A weak but statistically significant anti‑correlation is detected between the polarization spectral index, $\beta$, and weighted Faraday dispersion, $\sigma_{\rm RM,wtd}$, for two‑component systems, whereas one‑ and three‑component populations show no significant trend. The rest‑frame Faraday dispersion, $\sigma_{\rm RM,rest}$, exhibits significant positive correlations with redshift for the external‑screen dominated and mixed depolarization populations, even after controlling for radio luminosity, indicating increasingly turbulent or strongly magnetized environments surrounding radio AGN at earlier cosmic epochs.
\end{abstract}

\begin{keywords}
galaxies: magnetic fields -- polarization -- Faraday rotation -- intergalactic medium -- radio continuum: galaxies -- surveys -- active galactic nuclei -- intergalactic medium
\end{keywords}

\section{Introduction}

Polarized radio emission from extragalactic radio sources serves as a powerful diagnostic of magnetized plasma in and around active galactic nuclei (AGN). As synchrotron radiation travels through a magneto-ionic medium, its polarization characteristics are altered by Faraday rotation and depolarization processes, which depend on the electron density, magnetic field strength, and magnetic field geometry along the line of sight \citep{burn66, Sokoloff98}. The frequency-dependent polarization behaviour therefore carries important information about the physical conditions within the emitting source and its surrounding environment \citep{Purcell15, Ma19, Ma19b}. Broad-band radio spectro-polarimetry has therefore become an essential tool for investigating the magneto-ionic structure of radio galaxies and for studying cosmic magnetism on both galactic and extragalactic scales \citep{McClure_Griffiths10, Gaensler15, Costa16, Gaensler25}.

The development of modern radio facilities such as the Low-Frequency Array (LOFAR), the Karl G. Jansky Very Large Array (VLA), ASKAP, and MeerKAT has enabled polarization observations with unprecedented bandwidth, sensitivity, and spectral resolution \citep{Taylor09, Carretti13, Gaensler15, Anderson15, Pasetto18, Passeto21, Wang22, Pictor24, Taylor24}. In particular, broad-band observations densely sample the polarization signal as a function of wavelength squared, allowing precise measurements of Faraday rotation while largely eliminating the traditional $n\pi$ ambiguity associated with narrow-band observations \citep{Ma19}. These data also enable the application of advanced spectro-polarimetric analysis techniques such as RM-synthesis \citep{Brentjens05} and `QU-fitting' \citep{OSullivan_2009, Farnsworth11, OSullivan_2012, OSullivan_2017}, which can recover complex Faraday structures that are unresolved in conventional rotation measure analyses.

Broad-band polarization studies have shown that many radio AGN are Faraday complex, requiring multiple Faraday rotating and depolarizing components to reproduce the observed polarization spectra \citep{Anderson15, OSullivan_2017, Pasetto18, Paul26}. Such complexity is often interpreted as evidence for turbulent and inhomogeneous magneto-ionic environments associated with radio jets, lobes, host galaxies, or surrounding intra-group and intra-cluster media. Variations in depolarization behaviour can therefore provide important insights into the magnetic field geometry, thermal plasma distribution, and environmental conditions of radio sources \citep{burn66, Laing08, Heald15}. In particular, low-frequency observations are especially sensitive to Faraday depolarization because most depolarization mechanisms scale strongly with wavelength, typically as $\lambda^2$ or $\lambda^4$ \citep{burn66, Sokoloff98, Schnitzeler18}. Even relatively small fluctuations in electron density or magnetic field structure can therefore produce substantial depolarization at metre wavelengths, making low-frequency observations ideal for probing weak turbulence and complex magneto-ionic structure.

The LOFAR Two-Metre Sky Survey (LoTSS) has opened a new parameter space for low-frequency polarization studies by providing sensitive polarization measurements for thousands of extragalactic radio sources across large areas of the sky \citep{Shimwell22, lofar23, Shimwell26}. The homogeneous nature of the LoTSS polarization catalogues enables statistical investigations of depolarization and Faraday complexity across large radio AGN populations. Such studies are essential for understanding how polarization properties depend on source morphology, environment, redshift, and radio luminosity, and for connecting detailed spectro-polarimetric analyses of individual objects with the broader evolution of radio galaxies and cosmic magnetism \citep{Govoni04, Banfield11, Hales12, Beck13, Stil14}.

In this work, we present a depolarization analysis of polarized radio sources selected from the LoTSS DR2 catalogue. Using broad-band spectro-polarimetric modelling of the frequency-dependent Q, U Stokes parameters \citep{OSullivan_2012, OSullivan_2017, Sun20, Beck21}, we investigate the Faraday complexity and depolarization behaviour of the source population. The primary aim of this work is to characterize the Faraday complexity of polarized radio sources and to constrain the dominant magneto-ionic processes responsible for the observed depolarization behaviour at low radio frequencies. 
We further investigate the occurrence and characteristics of multi\textendash component Faraday structures, and the relative contribution of internal versus external Faraday screens to the overall depolarization. We examine the intrinsic polarization angle differences between RM component models, the physical characteristics of the magneto–ionic environments surrounding the sources, and the dependence of depolarization behaviour on redshift and radio luminosity. We also explore trends in depolarization with source size and assess how these trends may reflect different classes of radio sources and their surrounding environments.

The paper is organized as follows. In Section~\ref{sec:polFar}, we introduce the basic concepts of radio polarization and Faraday rotation. Section~\ref{sec:depolmodels} describes the depolarization models adopted in this work. In Section~\ref{sec:DR2source}, we present the LoTSS DR2 catalogue and the sample selection procedure. The main observational results are presented in Section~\ref{sec:results}. In Section~\ref{sec:weightedFaraday}, we investigate the polarization-weighted Faraday properties and depolarization behaviour of the sources. Section~\ref{sec:redepol} examines the redshift and radio luminosity evolution of Faraday depolarization and its environmental implications. Finally, in Section~\ref{sec:discuss}, we summarize the main conclusions of this work.

\section{Polarization and Faraday Rotation}
\label{sec:polFar}

Linearly polarized radio emission from galaxies is a direct signature of synchrotron radiation produced by relativistic electrons spiralling in magnetic fields. The observed polarization properties therefore provide a powerful diagnostic of magnetic field strength and geometry in radio jets and lobes, as well as information about the orientation and magneto-ionic environment of the source.

The polarization state of electromagnetic radiation is fully described by the Stokes parameters \(I, Q, U,\) and \(V\), which are defined in terms of measurable intensity combinations. For the present work, we focus on linear polarization, which can be represented as the complex quantity
\begin{equation}
P = Q + iU = p I e^{2i\psi},
\end{equation}
where \(p\) is the fractional linear polarization and \(\psi\) is the polarization angle. The fractional polarization is given by
\begin{equation}
p = \frac{P_{\rm pol}}{I} = \sqrt{q^2 + u^2},
\end{equation}
with \(P_{\rm pol}\) the polarized intensity, \(q = Q/I\), and \(u = U/I\). The polarization angle is then
\begin{equation}
\psi = \frac{1}{2}\arctan\left(\frac{U}{Q}\right),
\end{equation}
where the factor of \(1/2\) reflects the \(180^\circ\) ambiguity inherent in linear polarization.

A key physical process affecting polarized radio emission is Faraday rotation, in which the plane of polarization rotates as the wave propagates through a magnetized ionized medium. The accumulated rotation depends on the thermal electron density and the line-of-sight magnetic field, providing a direct probe of astrophysical magneto-ionic environments. This effect is commonly quantified using the Faraday depth,
\[
\varphi(s) = 0.81 \int_0^s n_e(s')\,B_\parallel(s')\,ds' \quad [\mathrm{rad\,m^{-2}}],
\]
where \(n_e\) denotes the thermal electron density in \(\mathrm{cm^{-3}}\), \(B_\parallel\) represents the component of the magnetic field aligned with the line of sight in \(\mu\mathrm{G}\) defined as positive when directed toward the observer, and \(s\) corresponds to the path length in parsecs, measured from the observer to the source \citep{Ferri21}.

In the simplest case - a single, uniform Faraday-rotating screen in front of a polarized emitter - the observed rotation measure (RM) equals the Faraday depth of that screen. Under this approximation, the observed polarization angle varies linearly with $\lambda^2$,

\begin{equation}
\psi(\lambda^2) = \psi_0 + \mathrm{RM}\,\lambda^2,
\end{equation}
where \(\psi_0\) represents the intrinsic polarization angle at the source \citep{Anderson15}. RM is the Faraday rotation measure, and the complete Faraday depth distribution encodes the structural complexity of the emitting and Faraday-rotating medium along the line of sight \citep{Brentjens05}.

Departures from the idealized single-screen behaviour arise when emission and Faraday rotation are co-spatial or when multiple magneto-ionic regions contribute along the line of sight. In such cases, different emitting regions experience different amounts of rotation, leading to wavelength-dependent cancellation of polarized emission \citep{Paul26}. This results in Faraday depolarization, in which the observed fractional polarization typically decreases with increasing wavelength due to differential rotation and turbulence in the magnetized plasma \citep{burn66, Sokoloff98, Laing08, Schnitzeler15, OSullivan_2017}.

Faraday depolarization therefore encodes information not only about ordered magnetic fields but also about fluctuations in electron density and magnetic field structure. Quantifying these effects requires broadband polarimetric modelling that goes beyond simple linear RM fitting, particularly at low radio frequencies where depolarization effects are strongly enhanced.

\section{Depolarization Models}
\label{sec:depolmodels}
Depolarization occurs because variations in the magnetic field or electron density along the line of sight, or across the telescope beam, cause the polarization angle to rotate by different amounts in different regions, reducing the net observed polarization \citep{burn66, Goodlet04}. When these diverse polarization orientations are averaged across the observing beam, they tend to cancel each other out, diminishing the overall polarization signal. Hence, two types of depolarization come into play: internal and external \citep{Pasetto18, Passeto21}. The remaining polarization percentage is effectively utilized to gauge the degree of order and orientation within the magnetic field, whether it resides in the radio source itself or in the intervening inter galactic medium (IGM). The complex polarized signal observed from radio sources is fundamentally described by \textit{Model M1} (baseline Faraday thin; \citealt{burn66}):
\begin{equation}
    p(\lambda^2;p_0;\psi_0;RM) = p_0 e^{2i(\psi_0+RM\lambda^2)} 
    \label{eq:M1}
\end{equation}
Here, \(p_0\) denotes the intrinsic fractional polarization of the emitted radiation, while \(\psi_0\) corresponds to the intrinsic polarization angle at the source. The quantity \(RM\lambda^2\) characterizes the Faraday rotation introduced by the intervening magneto--ionic medium, where RM is the rotation measure and \(\lambda^2\) represents the square of the observing wavelength.

In real scenarios, the magnetic field is not perfectly ordered, there exists random and uniform fields causing depolarization \citep{Goodlet04}. The various depolarization contributions are thus added to this fundamental equation \ref{eq:M1}. When the polarized radiation passes through an external magneto-ionic medium with a turbulent magnetic field, the depolarization caused is called `external Faraday dispersion'. This is modelled as \textit{Model M2}:
\begin{equation}
    p(\lambda^2;p_0;\psi_0;RM;\sigma_{RM}) = p_0 e^{2i(\psi_0+RM\lambda^2)} e^{-2\sigma^2_{RM}\lambda^4}
    \label{eq:M2}
\end{equation}
where \(\sigma_{RM}\) is the Faraday dispersion about the mean RM across the radio source on the sky \citep{burn66}.
The parameter $\sigma_{\rm RM}$ is commonly interpreted as representing random RM fluctuations arising in a Faraday-active medium external, but local, to the radio source \citep{Goodlet04, Laing08, OSullivan_2017}.

In a specific scenario, where the region producing the synchrotron radiation and the area causing the Faraday rotation are co-spatial, the resulting depolarization is internal, referred to as `differential Faraday rotation' \citep{Sokoloff98}. However, in this particular model, the magnetic field is assumed to be purely regular, meaning any turbulent magnetic field component is neglected. Under these conditions, radiation from different depths within the source will undergo varying amounts of Faraday rotation, as the path length through the magnetized medium differs for each region. This is modelled as a `Faraday thick' component, \textit{Model M3}:
\begin{equation}
    p(\lambda^2;p_0;\psi_0;RM;\Delta\mathrm{RM}) = p_0 e^{2i(\psi_0+RM\lambda^2)} \,\mathrm{sinc}(\Delta\mathrm{RM}\,\lambda^2 / \pi)
    \label{eq:M3}
\end{equation}
where \(\Delta\mathrm{RM}\) parametrizes the rotation depth (equivalent to \(R\) with observable \(RM = R/2\)). The parameter $\Delta{\rm RM}$ can account for both internal and external Faraday depolarization effects, such as a linear RM gradient across the emitting region or internal Faraday rotation within a uniform magnetic field \citep{Sokoloff98, Schnitzeler15}.

Furthermore, depolarization can also arise when there is a gradient in the Faraday depth \citep{OSullivan_2017} across the source itself, or within a foreground screen that is local to the emitting region, even if the magnetic field within that region is considered uniform. In such instances, the Faraday depth effectively takes the form of a Rotation Measure (RM) that varies across the telescope's observing beam. This spatial variation in RM causes different parts of the signal within the beam to have distinct polarization orientations, which then tend to cancel each other out when averaged, resulting in observed depolarization. A combined model incorporating both external dispersion and internal gradients, \textit{Model M4}, is:

\begin{equation}
\begin{aligned}
p(\lambda^2; p_0; \psi_0; RM; \sigma_{RM}; \Delta\mathrm{RM}) 
&= p_0 e^{2i(\psi_0 + RM\lambda^2)} \\
&\quad \times \mathrm{sinc}(\Delta\mathrm{RM}\,\lambda^2 / \pi)\,
e^{-2\sigma_{RM}^2\lambda^4}
\end{aligned}
\label{eq:M4}
\end{equation}

In general, depolarization at longer wavelengths can result from a combination of effects, including internal Faraday dispersion, when the synchrotron-emitting plasma and the magnetized, ionized medium responsible for Faraday rotation are mixed along the line of sight, as well as beam depolarization caused by the limited spatial resolution of our observations (for more details see \citet{Sokoloff98}).

To characterize the broadband polarization and Faraday rotation measure (RM) properties of our sources, we adopt the QU-fitting and model-selection framework introduced by \citep{OSullivan_2012, OSullivan_2017}. Combining the depolarization models (M1 - M4), we model a wide variety of Faraday rotation behaviours using a multi-component representation of the complex polarization (\(P\)), incorporating the influence of both ordered and turbulent magnetic fields \citep{Sokoloff98, OSullivan_2017}. The functional form adopted to describe the LoTSS-DR2 foreground structures associated with polarized background sources is based on the generalized formulation of \textit{Model M4}:

\begin{equation}
P_j = p_{0,j} \, I\, e^{2 i (\psi_{0_j}+{\rm RM}_j \lambda^2)}  \,\frac{\sin (\Delta {\rm RM}_j \lambda^2)}{\Delta {\rm RM}_j \lambda^2} \,e^{-2\sigma^2_{{\rm RM}_j} \lambda^4},
\label{modeleqn}
\end{equation}
where, for the \(j^{\mathrm{th}}\) component, \(p_{0,j}\) denotes the intrinsic fractional polarization, \(\psi_{0,j}\) is the intrinsic polarization angle, and \(\Delta {\rm RM}_j\) and \(\sigma_{{\rm RM}_j}\) characterize RM variations arising from ordered and turbulent magnetic fields, respectively (cf.~M3 and M2). Here, \(I\) represents the integrated flux density of the source. The linear polarization angle is defined conventionally as \(\psi = \frac{1}{2}\arctan\!\left(\frac{U}{Q}\right)\).

The LoTSS DR2 RM Grid \citep{lofar23} lacks reliable channelized Stokes I spectra\footnote{\url{https://github.com/sposullivan/LoTSS-RM-Grid/blob/main/example_RMtable_PolSpectra.ipynb}}, so we construct a model $I(\nu)$ from the catalogue 144 MHz flux density assuming a spectral index $\alpha=-0.7$ for all the catalogue sources.

To model more complicated behaviours than a single component, we use up to three RM components: $P = P_1 + P_2 + P_3$. For each source, we fit 1-, 2-, and 3-component combinations, systematically testing the four depolarization cases per component (M1--M4 equivalents): (i) Faraday thin (no $\sigma_\mathrm{RM}$, no $\Delta\mathrm{RM}$), (ii) $\sigma_\mathrm{RM}$ only, (iii) $\Delta\mathrm{RM}$ only, and (iv) both. This ensures we can accurately model complex foreground large scale structures of polarized background sources. In total, 12 models were fitted to each source. For every model, we computed both the reduced chi-squared statistic (\(\chi_r^2\)) and the Bayesian Information Criterion (BIC). The preferred model was then identified as the one with the minimum BIC value \citep{Raftery95}.

\subsection{The priors}
In QU-fitting, setting good starting assumptions (priors) for the model parameters is important because they help guide the fitting process in the right direction. The selection of these priors is often based on the observed Faraday spectrum of the source \citep{Pictor24, Piras25}. For example, when combining multiple models, The total initial polarization fraction $\sum p_0$ across all astrophysical components must satisfy $\sum p_0 \leq 1$. The instrumental leakage term is excluded from this physical constraint, and each initial polarization angle \(\psi_0\) is limited between 0 and 180 degrees for all the models considered here. 

Instrumental polarization, which often appears near Faraday depth 0 ${\rm rad \ m^{-2}}$ in LOFAR polarization observations, also requires specific prior constraints. Instrumental polarization can manifest across a range of RM values because it arises from RM fluctuations caused by the ionosphere throughout the observation \citep{Mevius18}. The instrumental polarization leakage is modelled as a Faraday-thin or top-hat Faraday thick (Burn-slab) component with its Rotation Measure confined near zero ($-3 \, {\rm rad \ m^{-2}}$ to $1.5 \, {\rm rad \ m^{-2}}$ for Faraday thin and $-6 \, {\rm rad \ m^{-2}}$ to $3.0 \, {\rm rad \ m^{-2}}$ for thick components respectively), capturing the typical RM range where leakage occurs in LOFAR data \citep{lofar23, Pirasthesis2024, Piras25, Paul26}. We adopt a uniform prior to constraint the instrumental polarization. This method allows for the separation of intrinsic source polarization from instrumental artifacts, enhancing the accuracy and reliability of the resulting Faraday rotation measures and polarization parameters. The value of $\sigma_{\mathrm{RM}}$ is constrained to be positive and below $1\,\mathrm{rad\,m^{-2}}$ in order to be detectable with LOFAR.

Moreover, for an external Faraday dispersion component, the fractional polarization scales as, \[
|p(\lambda^2)| = p_0 \exp\left(-2 \sigma_{\text{RM}}^2 \lambda^4\right), \] \citep{burn66}, so that even intrinsically 100\% polarized emission is exponentially suppressed once \( 2 \sigma_{\text{RM}}^2 \lambda^4 \gg 1 \). For the LOFAR HBA band (120–168 MHz, $\lambda^2 \approx 3.2 \, \rm{m}^2 \, - \, 6.3 \, \rm{m}^2$), this means that components with $\sigma_{\rm RM} \gtrsim 0.1 \, - \, 0.2 \,\, \text{rad} \, \text{m}^{-2}$ are already heavily depolarized ($|p|/p_0 \lesssim 10^{-2}$) across most of the band, and for $\sigma_{\rm RM} \gtrsim 0.3 \, - \, 0.5 \,\, \text{rad} \, \text{m}^{-2}$ they would be effectively undetectable at LoTSS sensitivity ($|p|/p_0 \ll 10^{-3}$) even if intrinsically strongly polarized.

\section{The LoTSS DR2 Catalogue and Sample selection}
\label{sec:DR2source}
The LOFAR Two-metre Sky Survey (LoTSS) Data Release 2 (DR2) provides a large catalogue of 2,461 extragalactic sources with high-precision Rotation Measure (RM) measurements across 5,720 square degrees of the sky \citep{lofar23}. The RM Grid catalogue, together with the corresponding Stokes \(Q\) and \(U\) frequency spectra at 20\arcsec\ resolution for each source, is publicly available at \url{https://lofar-mksp.org/data/}. This dataset achieves a polarized source density of approximately 0.43 sources per square degree and was generated by deriving linear polarization and RM properties through RM synthesis applied to Stokes \(Q\) and \(U\) channel images. The observations span the frequency range 120 - 168 MHz with a channel width of 97.6 kHz and were obtained at an angular resolution of 20\arcsec. Approximately \(0.2\%\) of total-intensity sources brighter than 1 mJy beam\(^{-1}\) were detected in polarization, with a median detection limit of 0.6 mJy beam\(^{-1}\). The catalogue reports a median RM uncertainty of 0.06 rad m\(^{-2}\), although residual systematic uncertainties of up to 0.3 rad m\(^{-2}\) may persist after ionospheric correction \citep{lofar23}. 

The polarization SNR is computed as the median of \( |P|/|P|_{\mathrm{err}} \) across all valid data points, where \( |P| = \sqrt{Q^2 + U^2} \) is the observed polarization amplitude and \( |P|_{\mathrm{err}} = \sqrt{Q_{\mathrm{err}}^2 + U_{\mathrm{err}}^2} \) is its associated uncertainty, including only points with \( |P|_{\mathrm{err}} > 0 \). The use of the median SNR as a selection criterion provides a robust measure of detection significance that is relatively insensitive to outliers arising from individual noisy measurements.

A total of 1744 unique polarized sources were retained in the extended sample by requiring a median channel polarization signal-to-noise ratio of \({\rm SNR} \geq 1\). This inclusive threshold was adopted to maximize sensitivity to weakly polarized and potentially strongly depolarized systems, while still excluding spectra dominated entirely by noise. Since low-frequency depolarization can significantly suppress the observed polarized intensity, imposing a stricter polarization threshold at the selection stage could bias the sample against intrinsically Faraday-complex sources. For each pre-selected source, we additionally required at least 5 valid wavelength-squared points after finite masking, at least one point with positive polarization error, successful convergence of at least one of the 12 models, and reduced chi-squared below 3.0 for the lowest-BIC model. These conservative thresholds ensure robust multi-component depolarization modelling via broad-band spectro-polarimetric `QU-fitting’ approach \citep{Farnsworth11, OSullivan_2012, Anderson15}. A total of 1565 sources satisfy the pre-selection and post-fit quality control. Only those sources meeting all of the following criteria have been analyzed and tabulated in Tables~\ref{table:modelpara}.

In this work, we model only the polarization from the single $20\arcsec$ beam at the polarized peak, which primarily traces the core and inner-jet regions rather than the extended, low-surface-brightness lobes. From visual inspection, we find that only \(\sim 10-15\%\) of the source spectra can be associated with extended lobe emission, while the remainder are predominantly core/jet dominated. Considering only sources with available redshift and size measurements, we find that $\sim54\%$ of the analysed sample are spatially resolved, while the remaining sources are unresolved at the $20\arcsec$ LoTSS beam resolution. For these sources, the median projected linear size is $\sim461$ kpc, corresponding to a median angular size of $\sim74\arcsec$.

The $Q(\lambda^{2})$ and $U(\lambda^{2})$ spectra were extracted from the LoTSS-DR2 RM Grid catalogue at the single pixel corresponding to the peak polarized flux density position of each source. The LoTSS-DR2 RM Grid catalogue was derived from the LoTSS DR2 Stokes $Q$ and $U$ frequency cubes over the frequency range 120--168~MHz at an angular resolution of $20^{\prime\prime}$.

Since a large fraction of the sources are spatially resolved at this resolution, the extracted spectra do not represent the integrated polarization properties of the full radio source. Instead, they characterize the beam-averaged polarization behaviour local to the polarized peak region, which may correspond to compact hotspots, jet features, or localized lobe structures. Consequently, the derived Faraday properties primarily probe the magneto-ionic environment and depolarization behaviour within the sampled polarized region rather than the source as a whole.

\begin{figure}
    \centering
    \includegraphics[width=0.95\linewidth]{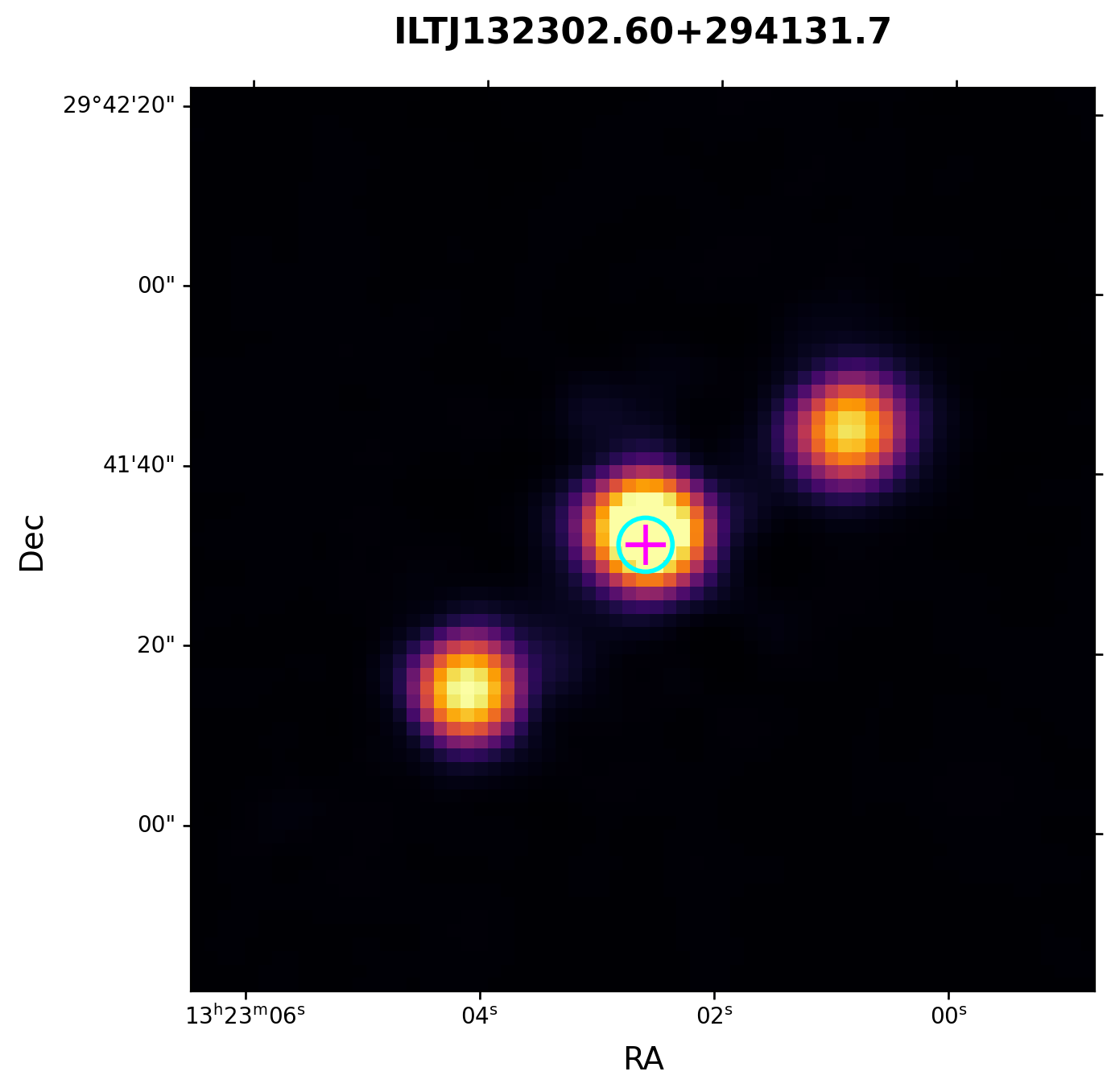}
   \caption{LoTSS-DR3 \(6^{\prime\prime}\)-resolution image of one of the analysed polarized radio sources. The magenta `+' symbol marks the location of the polarized detection from the LoTSS--DR2 RM Grid catalogue, while the cyan circle indicates the \(20^{\prime\prime}\) synthesized beam used for the extraction of the \(q(\lambda^2)\) and \(u(\lambda^2)\) spectra. Since many sources in the sample are spatially resolved at \(20^{\prime\prime}\) resolution, the extracted polarization spectra probe the local magneto-ionic properties within the beam centred on the polarized peak rather than the integrated properties of the entire radio source.}
   \label{fig:img_LoTSS20}
\end{figure}

Figure~\ref{fig:img_LoTSS20} presents example LoTSS images\footnote{\url{https://github.com/mhardcastle/lotss-cutout-api}} of the analysed source at angular resolutions of \(6^{\prime\prime}\) \citep{Shimwell26}. In all images, the cyan circle indicates the \(20^{\prime\prime}\) resolution beam at the location of the polarized detection, marked by the magenta `+' symbol. The full set of analysed source images is available through the links provided in the Data Availability section.

\section{Results}
\label{sec:results}

Table~\ref{tab:fri_pol_fits} presents the best-fitting depolarization
model parameters for all LoTSS-DR2 sources, including the intrinsic
instrumental polarization fraction ($P_0^{\rm inst}$), astrophysical
intrinsic polarized fractions ($P_0$), rotation measures (RM), Faraday
dispersion terms ($\sigma_{\rm RM}$), differential Faraday rotation
terms ($\Delta{\rm RM}$), instrumental RM contributions
(RM$_{\rm inst}$ and $R_{\rm inst}$), intrinsic polarization angles
($\psi_0^{\rm inst}$ and $\psi_0$), together with the reduced
$\chi_r^2$ and Bayesian Information Criterion (BIC) values.

The best-fitting models provide an accurate representation of the broadband polarization behaviour, with $\chi_r^2$ values ranging from 0.86 to
2.97 and a median value of 1.51. Out of 1565 sources, 56.8\% are
best described by a single Faraday component, 25.1\% require two
components, and 18.1\% require three components. Within the limits
of our measurements, 16.2\% of sources are classified as Faraday
thin, showing neither measurable Faraday dispersion
($\sigma_{\rm RM}=0$) nor differential Faraday rotation
($\Delta{\rm RM}=0$). A total of 75.7\% of the sample require one or
more non-zero $\sigma_{\rm RM}$ components, while 20.0\% require
non-zero $\Delta{\rm RM}$ terms. Additionally, 11.9\% of sources
require a combination of both $\sigma_{\rm RM}$ and
$\Delta{\rm RM}$ components in the best-fitting model. Furthermore,
9.8\% of the sample show intrinsic instrumental polarization fractions
($P_0^{\rm inst}$) exceeding the summed astrophysical polarized
fraction ($\sum P_0$), suggesting that instrumental leakage or
calibration residuals may contribute significantly to the observed
polarization signal in these systems. The fitted instrumental rotation measures span a range from $-3.00$ to $1.50~{\rm rad~m^{-2}}$, with a median value of $-0.22~{\rm rad~m^{-2}}$.

\begin{table*}
\centering
\scriptsize
\setlength{\tabcolsep}{5pt}
\renewcommand{\arraystretch}{1.0}

\caption{Sample columns of the polarization model fit parameters, with $1-\sigma$ uncertainties given in parentheses, are shown for LoTSS-DR2 sources.}
\label{tab:fri_pol_fits}

\begin{tabular}{@{}p{2.3cm}p{0.75cm}p{1.65cm}p{1.65cm}p{1.5cm}p{1.5cm}p{0.75cm}p{0.75cm}p{1.0cm}p{1.65cm}p{0.65cm}p{0.65cm}@{}}

\toprule
Source Name DR2 & $P_0^{\rm inst}$ & $P_0$(1,2,3) &
RM(1,2,3) & $\sigma_{\rm RM}$(1,2,3) &
$\Delta{\rm RM}$(1,2,3) &
RM$_{\rm inst}$ & $R_{\rm inst}$ &
$\psi_0^{\rm inst}$ &
$\psi_0$(1,2,3) &
$\chi_r^2$ & BIC \\
 & (\%) & (\%) & (rad m$^{-2}$) &
(rad m$^{-2}$) & (rad m$^{-2}$) &
(rad m$^{-2}$) & (rad m$^{-2}$) &
(deg) & (deg) & & \\
\midrule

ILTJ000506.83+405711.8 & 0.2(0.1) & 10.7(0.5), -, - &
-62.96(0.01), -, - & 0.15(0), -, - &
-, -, - & -0.41(0.23) & - &
90(60.6) & -39.5(2.7), -, - &
1.22 & 1132.2 \\

ILTJ000559.68+350204.4 & 2.3(0.7) & 1.3(0.8), -, - &
-7.84(0.28), -, - & -, -, - &
0.67(0.08), -, - & - &
-0.6(0.04) & 8.9(7) &
22.1(70.2), -, - &
2.62 & 2389.9 \\

ILTJ000624.41+263545.6 & 0.1(0) &
3.4(0.1), 48.3(0.6), - &
-111.16(0.01), -83.08(4.51), - &
0.1(0), 0.53(0.05), - &
-, -, - & -0.1(0.13) & - &
77.3(35.9) &
-65.1(2.4), 30.6(876), - &
1.14 & 1084.5 \\

ILTJ000831.36+421725.0 & 0.4(0.1) &
100(468.1), -, - &
-5.85(1.4), -, - &
0.49(0.21), -, - &
-, -, - & -0.4(0.1) & - &
1.1(26.7) &
-20.6(274), -, - &
2.17 & 1968.8 \\

ILTJ001006.00+411442.5 & 0.4(0) &
1.4(0.1), -, - &
-63.77(0.01), -, - &
0.04(0.01), -, - &
-, -, - & -0.4(0.04) & - &
34.1(9.5) &
68.5(2.9), -, - &
1.58 & 1450.7 \\

ILTJ001007.40+304524.3 & 0.1(0.1) &
8.3(1.9), 44.1(10.7), 3.5(3.6) &
-64.05(0.04), -27.61(1.41), -0.6(0.6) &
0.24(0.01), 0.48(0.02), 0.26(0.05) &
-, -, - & -0.21(1.49) & - &
-90(507.3) &
21.8(8.3), -25.1(275.1), 17.3(110.7) &
1.19 & 1159.0 \\

ILTJ001025.52+332942.4 & 0(0) &
0.2(0.1), -, - &
-1.34(0.14), -, - &
0.09(0.09), -, - &
-, -, - & -0.03(0.77) & - &
90(209.5) &
-32.1(36.9), -, - &
2.97 & 2702.5 \\

ILTJ001028.84+204748.4 & 0.2(0.1) &
3.9(1.2), 48(14.7), 0.1(0.2) &
-32.33(0.03), -13.96(7.93), 7.44(0.38) &
0.13(0.01), 0.51(0.09), 0(2966.1) &
-, -, - & -0.6(0.3) & - &
-90(82.7) &
15.5(7.1), 76.6(1551.1), -50.1(104.5) &
0.96 & 935.8 \\

ILTJ001247.35+335338.7 & 0.6(0.3) &
0.6(0.3), -, - &
-10.6(0.29), -, - &
-, -, - &
-, -, - & 0.25(0.3) & - &
-90(83.4) &
-34.3(78.6), -, - &
1.57 & 1326.4 \\

ILTJ001355.74+422428.9 & 49.7(0.1) &
0.1(0), 50(9.4), - &
-19.63(0.46), -0.44(0.21), - &
-, -, - &
-, -, - & -0.44(0.22) & - &
-27.9(62) &
-43.5(120.4), 61.6(61.7), - &
2.53 & 2326.1 \\

ILTJ001356.36+191041.1 & 1(0.4) &
5(1.1), 47.5(4.3), - &
-16.82(0.06), 4.1(1.51), - &
0.08(0.04), 0.39(0.03), - &
-, -, - & -2.34(0.23) & - &
-90(63.2) &
-5.2(16.4), -6.4(300.2), - &
1.87 & 1737.6 \\

ILTJ001454.48+222611.6 & 1(0.5) &
4(4.3), 48(19.9), - &
-12.04(0.36), 13.23(0.41), - &
0.18(0.11), 0.32(0.02), - &
-, -, - & -0.21(0.32) & - &
90(86.3) &
31.2(82.9), 13.3(85.2), - &
2.83 & 2577.5 \\

ILTJ001536.10+305223.9 & 0.2(0.1) &
7.9(0.2), 46.1(0.6), - &
-63.17(0.01), -48.41(0.58), - &
0.12(0.01), 0.39(0.02), - &
0.48(0.01), 0.86(0.03), - &
- & -0.27(0.15) &
32.5(3.7) &
20.8(1.6), 63.2(115.2), - &
1.20 & 1150.3 \\

ILTJ001612.28+323857.4 & 0.1(0) &
1.8(0.1), 49.1(1.3), - &
-49.91(0.03), -24.4(2.8), - &
0.2(0.01), 0.53(0.02), - &
-, -, - & -0.1(0.05) & - &
90(12.9) &
26.1(6), -11.7(561), - &
1.40 & 1209.9 \\

ILTJ001641.82+313901.6 & 0.2(0.1) &
0.3(0.1), -, - &
-8.47(0.27), -, - &
-, -, - &
-, -, - & 0.43(0.38) & - &
-90(103.2) &
-88.5(72.4), -, - &
1.41 & 1303.0 \\

\bottomrule
\end{tabular}

\vspace{0.3cm}

\parbox{\textwidth}{
\footnotesize
\textbf{Note.}
A sample of 15 sources from the full LoTSS-DR2 depolarization
catalogue is shown here for guidance regarding the table format and
best-fitting model parameters. Entries marked with `-' indicate that
the corresponding parameter is not required by the best-fitting model.
The complete catalogue containing all 1565 sources is available
through the links provided in the Data Availability section.
}
\label{table:modelpara}
\end{table*}

\begin{figure}
    \centering
    \includegraphics[width=0.9\linewidth]{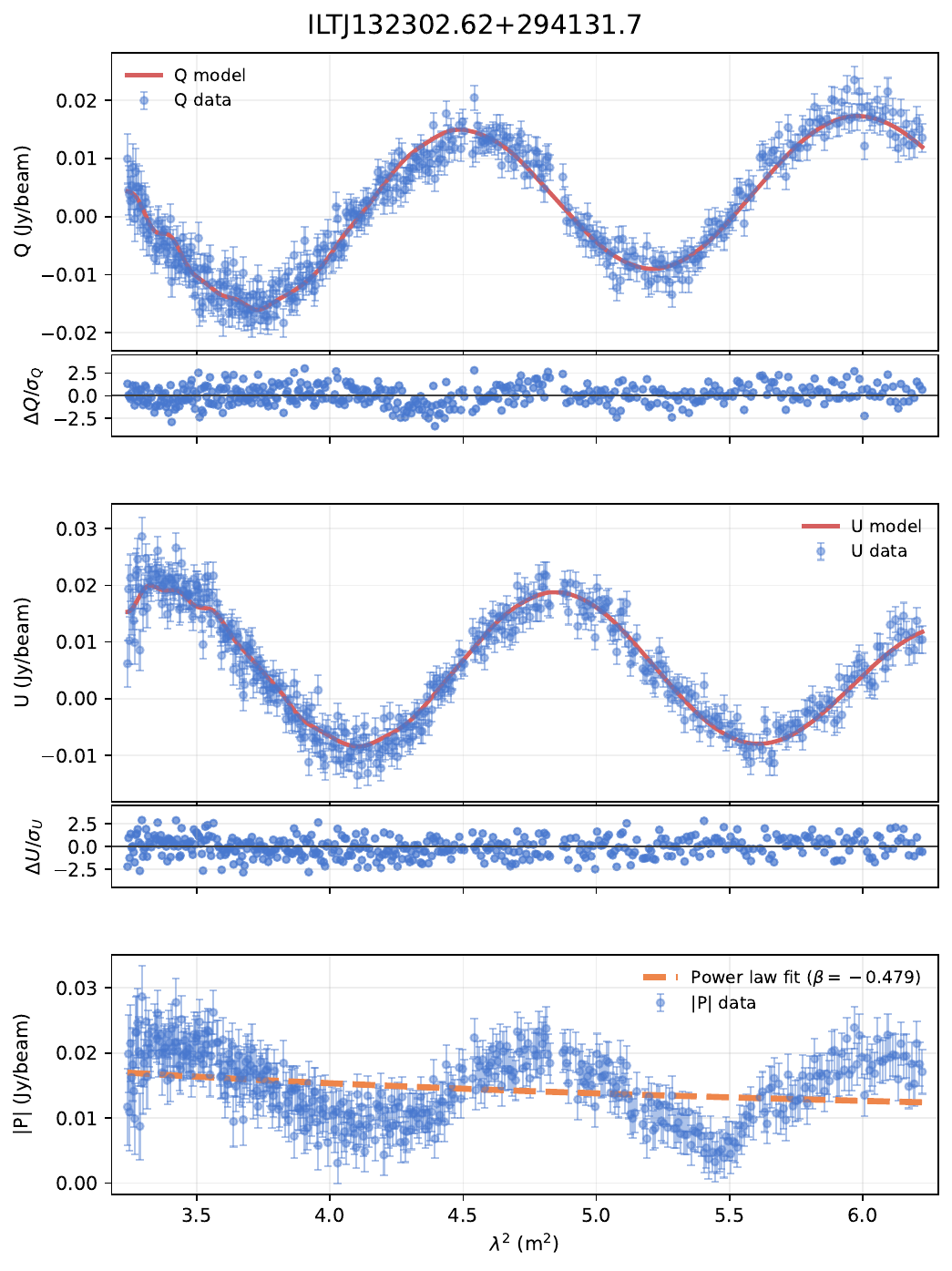}
    \caption{Example of the best-fitting polarization model for the source ILTJ132302.62+294131.7. The top two panels show the observed Stokes \(Q\) and \(U\) spectra as a function of \(\lambda^2\), where the light-blue points with error bars represent the LOFAR HBA measurements and the red curves show the best-fitting model. The sub-panels display the normalized residuals, \(\Delta Q/\sigma_Q\) and \(\Delta U/\sigma_U\), respectively. The higher polarization signal-to-noise ratio (\({\rm SNR}\approx4.5\)) allows the oscillatory Faraday structure in the \(Q\) and \(U\) spectra to be clearly resolved. Despite the slightly elevated reduced chi-square value (\(\chi^2_{\rm r}\approx1.2\)), the residuals show no strong frequency-dependent structure, indicating that the dominant Faraday behaviour is well reproduced by the model. The bottom panel shows the polarized intensity, \(|P|=\sqrt{Q^2+U^2}\), as a function of \(\lambda^2\). The dashed orange curve indicates the best-fitting power-law model, \(P=A(\lambda^2)^\beta\), with \(\beta=-0.48\), consistent with moderate depolarization across the LOFAR HBA band.}
   \label{fig:modelfit}
\end{figure}

Figure~\ref{fig:modelfit} shows the polarization spectrum and best-fitting Faraday model for the source ILTJ132302.62+294131.7. Owing to its relatively high polarization signal-to-noise ratio (\({\rm SNR}\approx4.5\)), the frequency-dependent structure in the Stokes \(Q\) and \(U\) spectra is clearly resolved across the LOFAR HBA band. The best-fitting model consists of three external Faraday-dispersion components together with an additional instrumental polarization term.

The polarization spectrum is dominated by a strongly polarized component with intrinsic fractional polarization \(p_{0,3}\simeq0.49\), located at \({\rm RM}\simeq24.5~{\rm rad\,m^{-2}}\) and characterized by substantial Faraday dispersion (\(\sigma_{\rm RM}\simeq0.54~{\rm rad\,m^{-2}}\)). At LoTSS wavelengths, this level of Faraday dispersion produces strong wavelength-dependent depolarization, significantly attenuating the polarized emission toward longer wavelengths. A second component at \({\rm RM}\simeq2.1~{\rm rad\,m^{-2}}\), consistent with the catalogue LOFAR RM value, exhibits considerably weaker dispersion (\(\sigma_{\rm RM}\simeq0.09~{\rm rad\,m^{-2}}\)) and therefore remains visible across most of the observing band. A third, weaker component at negative RM contributes only marginally to the total polarized signal and is comparatively poorly constrained.

The higher SNR of this source makes the oscillatory structure in \(Q\) and \(U\) especially apparent, demonstrating the interference between multiple Faraday components with different rotation measures and depolarization scales. The residual panels, showing \(\Delta Q/\sigma_Q\) and \(\Delta U/\sigma_U\), remain largely structureless despite the slightly elevated reduced chi-square value (\(\chi^2_{\rm r}\approx1.2\)), indicating that the model captures the dominant Faraday behaviour without strong evidence for additional unresolved structure.

The bottom panel displays the polarized intensity, \(|P|=\sqrt{Q^2+U^2}\), as a function of \(\lambda^2\), together with the corresponding best-fitting power-law model,
\[
P = A\,(\lambda^2)^{\beta}.
\]
The fitted slope, \(\beta=-0.48\pm0.09\), indicates moderate depolarization across the LOFAR HBA band. Although the polarized intensity spectrum displays a relatively smooth monotonic decline with wavelength, the corresponding \(Q\) and \(U\) spectra reveal substantially richer Faraday complexity, highlighting the importance of full spectro-polarimetric modelling when interpreting low-frequency polarization data.

\subsection{Polarization-Weighted Faraday Properties and Depolarization}
\label{sec:weightedFaraday}
To enable a consistent comparison of RM dispersion among sources, we define the polarization-weighted RM dispersion as
$\sigma_{\mathrm{RM,wtd}} =
\frac{\sum_j p_{0,j}\,\sigma_{\mathrm{RM},j}}
{\sum_j p_{0,j}}$,
where \(p_{0,j}\) denotes the intrinsic polarized intensity of the \(j^{\mathrm{th}}\) RM component and \(\sigma_{\mathrm{RM},j}\) is the associated RM dispersion.
The associated uncertainty is calculated using standard propagation of independent errors as $\delta \sigma_{\mathrm{RM,wtd}} = \frac{\sqrt{\sum_j p_{0,j}^2 \, (\delta \sigma_{\mathrm{RM},j})^2}}{\sum_j p_{0,j}}$. Similarly, the polarization-weighted rotation measure is defined as $\mathrm{RM}_{\mathrm{wtd}} = \frac{\sum_j p_{0,j}\,\mathrm{RM}_j}{\sum_j p_{0,j}}$, with uncertainty $\delta \mathrm{RM}_{\mathrm{wtd}} = \frac{\sqrt{\sum_j p_{0,j}^2 \, (\delta \mathrm{RM}_j)^2}}{\sum_j p_{0,j}}$. We also calculate the polarization-weighted RM gradient as $\Delta\mathrm{RM}_{\mathrm{wtd}} = \frac{\sum_j p_{0,j}\,\Delta\mathrm{RM}_j}{\sum_j p_{0,j}}$, with uncertainty $\delta \Delta\mathrm{RM}_{\mathrm{wtd}} = \frac{\sqrt{\sum_j p_{0,j}^2 \, (\delta \Delta\mathrm{RM}_j)^2}}{\sum_j p_{0,j}}$. 

\begin{table*}
\centering
\scriptsize
\setlength{\tabcolsep}{4pt}
\renewcommand{\arraystretch}{1.0}
\caption{Basic depolarization properties of analysed sources in the sample.}
\label{tab:basic_depol}
\begin{tabular}{@{}p{2.5cm}p{1.1cm}p{1.1cm}p{1.2cm}p{1.0cm}p{1.3cm}p{1.5cm}p{1.4cm}p{1.4cm}p{0.9cm}p{1.1cm}p{1.1cm}@{}}
\toprule
Source Name DR2 & RA & DEC & $I_{144}$ & $z_{\rm best}$ & $<p_0>$ & ${\rm RM}_{\rm wtd}$ & $\sigma_{\rm RM,wtd}$ & $\Delta {\rm RM}_{\rm wtd}$ & $\beta$ & Linear Size & Angular Size \\
& (deg) & (deg) & (Jy beam$^{-1}$) & & (\%) & (rad m$^{-2}$) & (rad m$^{-2}$) & (rad m$^{-2}$) & & (kpc) & (arcsec) \\
\midrule
ILTJ000506.83+405711.8 & 1.27845 & 40.95328 & 0.206(0) & - & 10.71(0.49) & -62.96(0.01) & 0.15(0) & - & -1.03 & - & - \\
ILTJ000559.68+350204.4 & 1.49866 & 35.03456 & 0.138(0) & - & 1.3(0.8) & -7.84(0.28) & - & 0.67(0.08) & -0.85 & - & - \\
ILTJ000624.41+263545.6 & 1.60172 & 26.59602 & 0.774(0) & 0.811 & 45.36(0.56) & -84.92(4.22) & 0.5(0.04) & - & -0.46 & 1907.4 & 252.8 \\
ILTJ001006.00+411442.5 & 2.52502 & 41.24516 & 0.721(0) & - & 1.35(0.07) & -63.77(0.01) & 0.04(0.01) & - & 0.16 & - & - \\
ILTJ001007.40+304524.3 & 2.53082 & 30.75676 & 0.314(0) & - & 36.18(8.41) & -31.33(1.11) & 0.43(0.02) & - & -1.66 & - & - \\
ILTJ001025.52+332942.4 & 2.60634 & 33.49511 & 0.916(0) & 0.742 & 0.17(0.13) & -1.34(0.14) & 0.09(0.09) & - & -0.67 & 582.1 & 79.6 \\
ILTJ001028.84+204748.4 & 2.62016 & 20.79678 & 0.148(0.001) & 0.6 & 44.58(13.57) & -15.27(7.32) & 0.48(7.61) & - & -0.66 & - & - \\
ILTJ001247.35+335338.7 & 3.19728 & 33.8941 & 0.052(0) & 1.682 & 0.61(0.28) & -10.6(0.29) & - & - & -0.32 & - & - \\
ILTJ001355.74+422428.9 & 3.48225 & 42.40805 & 0.36(0) & - & 49.89(9.37) & -0.47(0.21) & - & - & -0.27 & - & - \\
ILTJ001356.36+191041.1 & 3.48484 & 19.17808 & 0.058(0) & 0.477 & 43.49(3.91) & 2.12(1.37) & 0.36(0.03) & - & -0.53 & - & - \\
ILTJ001454.48+222611.6 & 3.727 & 22.43656 & 0.052(0) & 0.955 & 44.58(18.33) & 11.27(0.38) & 0.31(0.02) & - & -0.7 & 1503.9 & 189.9 \\
ILTJ001536.10+305223.9 & 3.90043 & 30.87332 & 0.762(0.001) & - & 40.5(0.55) & -50.56(0.5) & 0.35(0.02) & 0.81(0.02) & -2 & - & - \\
ILTJ001612.28+323857.4 & 4.05116 & 32.6493 & 1.341(0.001) & 0.839 & 47.37(1.27) & -25.33(2.7) & 0.52(0.02) & - & -1.29 & 431 & 56.5 \\
ILTJ001641.82+313901.6 & 4.17424 & 31.65047 & 0.103(0) & 0.646 & 0.27(0.12) & -8.47(0.27) & - & - & -0.46 & 582.4 & 84.3 \\
ILTJ001808.57+310234.6 & 4.53572 & 31.04296 & 0.147(0) & 0.904 & 5.1(0.37) & -77.16(0.02) & 0.1(0.01) & - & -0.37 & 1632.3 & 209.3 \\
\bottomrule
\end{tabular}
\vspace{1mm}
\begin{minipage}{\textwidth}
\footnotesize{\textbf{Note.} A sample of basic depolarization properties of 15 sources from the full LoTSS-DR2 catalogue. The complete catalogue containing all 1565 sources is available through the links provided in the Data Availability section. Source Name DR2 is the LoTSS-DR2 source identifier; RA and DEC are the right ascension and declination in degrees; $I_{144}$ is the total intensity at 144 MHz in units of Jy beam$^{-1}$; $z_{\rm best}$ is the best available redshift estimate; $p_0$ is the polarization-weighted intrinsic fractional polarization in percent; ${\rm RM}_{\rm wtd}$ is the polarization-weighted Faraday rotation measure in units of rad m$^{-2}$; $\sigma_{\rm RM,wtd}$ is the polarization-weighted Faraday dispersion in units of rad m$^{-2}$; $\Delta {\rm RM}_{\rm wtd}$ is the polarization-weighted Faraday depth separation or RM gradient in units of rad m$^{-2}$; $\beta$ is the power-law index obtained from fitting the polarized intensity spectrum using $P = A(\lambda^2)^{\beta}$; Linear Size is the projected physical source size in kpc; and Angular Size is the apparent source size in arcsec.}
\end{minipage}
\end{table*}

For our full sample of 1565 polarized sources, the median intrinsic fractional polarization is 10.81\%, the median values of $\sigma_{\rm RM,wtd}$, and $\Delta {\rm RM}_{\rm wtd}$, are 0.3~rad~m$^{-2}$, 0.5~rad~m$^{-2}$, respectively. The median total intensity at 144~MHz is 0.150~Jy~beam$^{-1}$, while the median redshift of the sample is 0.61. Sources that are spatially resolved at the LoTSS resolution (i.e. angular sizes $>20$~arcsec; 54 \% of the sample) have a median angular size of 74.2~arcsec and a median linear size of 460.9~kpc. The median power-law index describing the polarized intensity spectrum, defined through $P = A(\lambda^2)^{\beta}$, is 
$\beta_{\rm med}=-0.44$, with values ranging from $-2.00$ to $1.97$. These results indicate that the polarized source population exhibits a broad range of depolarization behaviours across the LOFAR HBA band (Table \ref{tab:basic_depol}).

\subsection{Intrinsic Polarization Angle and RM Separations between Faraday Components}
We investigated the intrinsic polarization angle differences between Faraday components in the best-fitting multi-component RM models (Table~\ref{table:modelpara}) \citep{OSullivan_2017}. For sources with multiple RM components, we computed the pairwise intrinsic polarization angle differences, $|\psi_{0,i}-\psi_{0,j}|$, wrapped to the range $0^\circ$--$90^\circ$ to account for the $180^\circ$ ambiguity of linear polarization angles.

\begin{figure}
    \centering
    \includegraphics[width=0.49\linewidth]{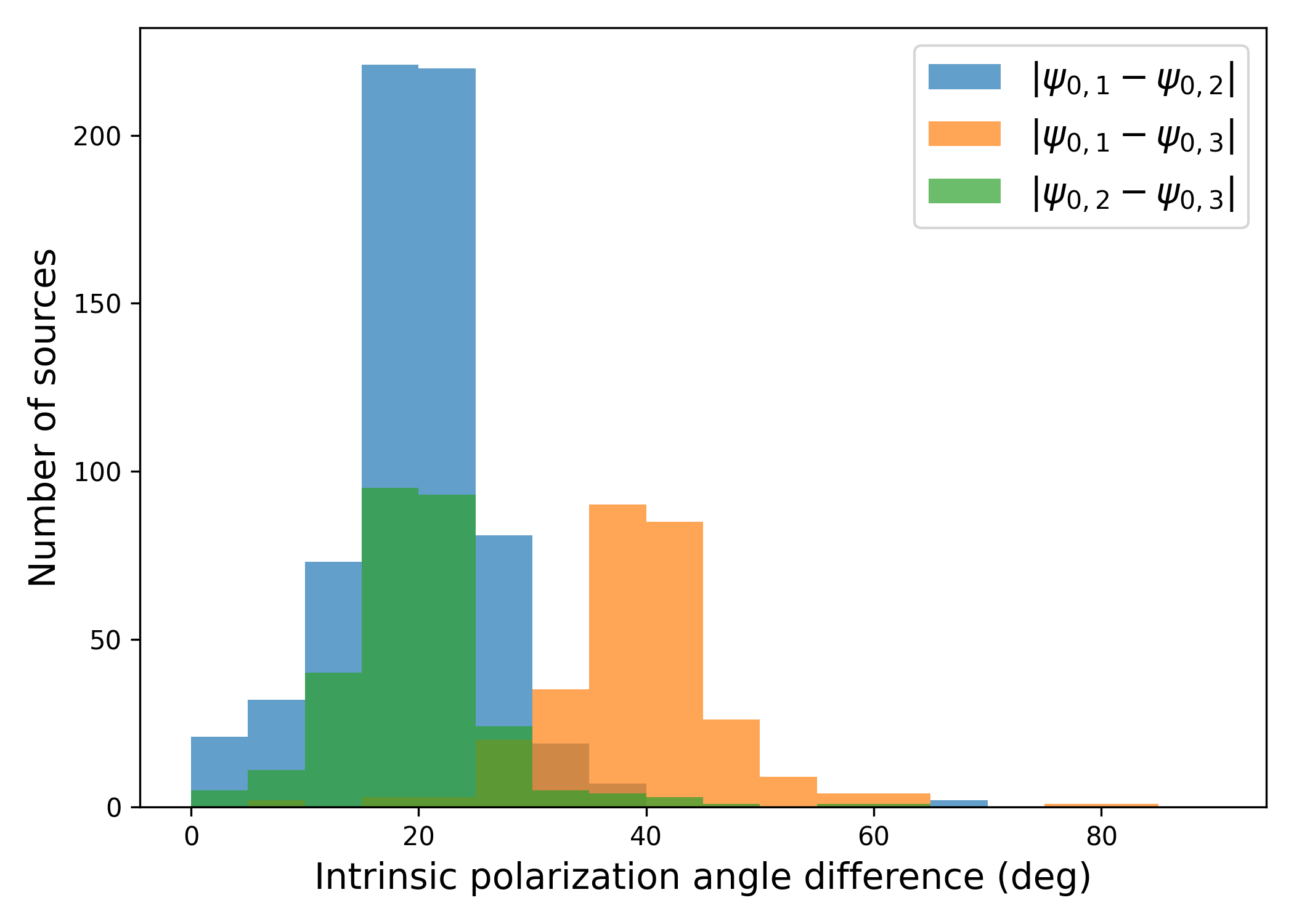}
    \includegraphics[width=0.49\linewidth]{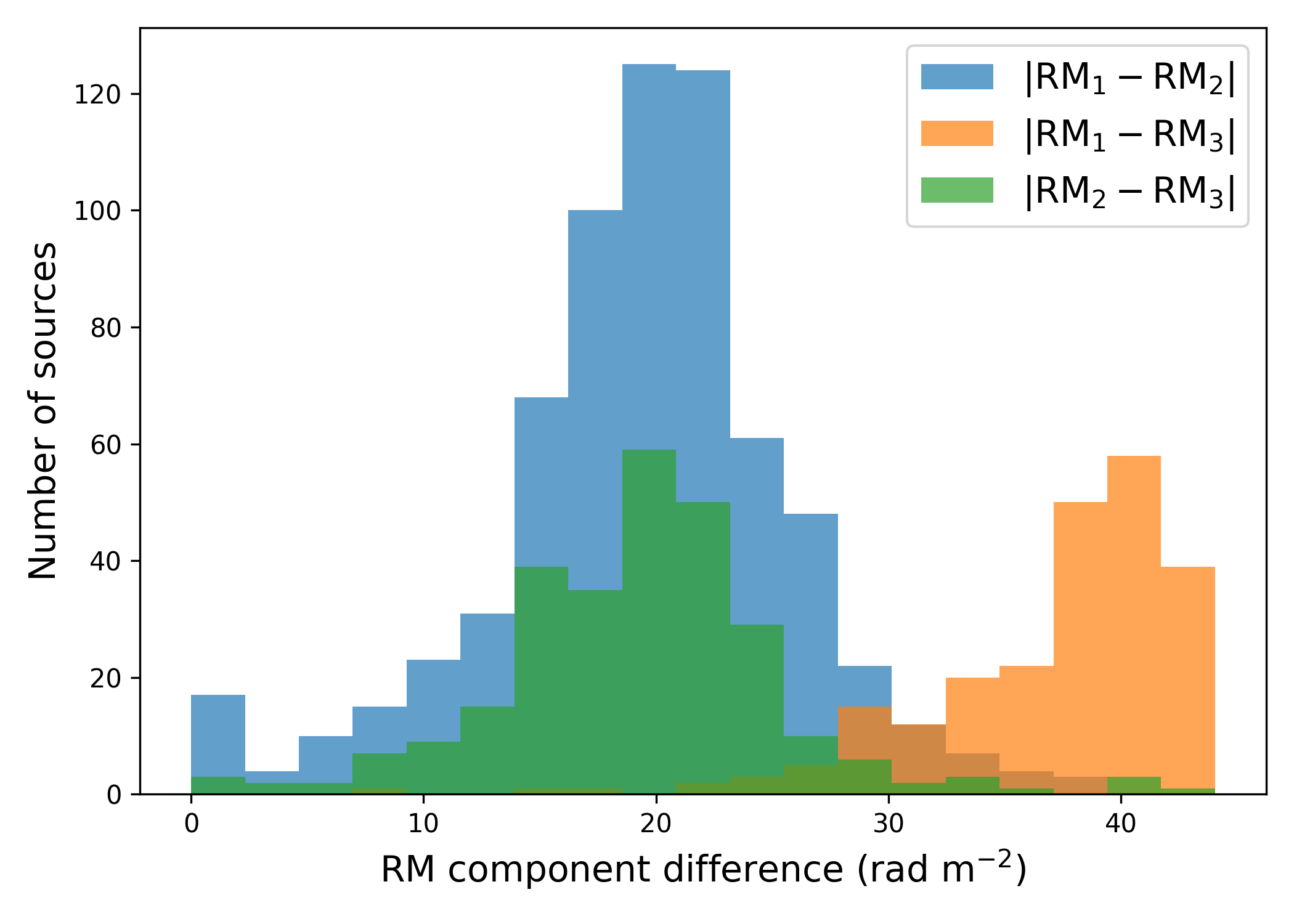}
    \caption{Pairwise intrinsic polarization angle and RM differences between components in the best-fitting multi-component polarization models. \textit{left:} Distribution of intrinsic polarization angle differences, wrapped to the range $0^\circ$--$90^\circ$ to account for the $180^\circ$ ambiguity of linear polarization angles. \textit{right:} Distribution of pairwise RM separations between fitted Faraday components. In both panels, blue, orange, and green histograms correspond to the component pairs $(1,2)$, $(1,3)$, and $(2,3)$, respectively. The distributions are strongly concentrated toward relatively small intrinsic polarization angle and RM separations, indicating that many of the fitted RM components are physically related and likely trace emission regions embedded within a common or smoothly varying magneto-ionic environment rather than independent orthogonal polarization structures.}
    \label{fig:psi0_RM_diff}
\end{figure}

Out of the total sample of 1565 sources, 393 ($25.1\%$) are best described by two RM components, while 283 ($18.1\%$) require three RM components. The distribution of intrinsic polarization angle differences is shown in Fig.~\ref{fig:psi0_RM_diff}. For two-component models, the distribution is strongly concentrated at relatively small angle separations, with a median value of $19.8^\circ$ and a mean of $19.5^\circ$. The pairwise angle differences in three-component models are somewhat broader, with a median separation of $28.9^\circ$, although they still predominantly occupy the low-angle regime.

Our sample shows no evidence for such a bi-modality or orthogonal preference \citep{OSullivan_2017}. Instead, the majority of RM components exhibit relatively aligned intrinsic polarization angles, typically differing by $\lesssim30^\circ$. This suggests that the polarized emission components in our sources are generally associated with magnetic field structures that are similarly oriented on the sky, rather than arising from independent orthogonal polarization regions. 

The predominance of small intrinsic polarization angle separations is accompanied by relatively small RM separations between fitted components. The RM difference distributions peak at $\sim 20\ {\rm rad\ m^{-2}}$, with median pairwise separations of $19.7\ {\rm rad\ m^{-2}}$ for two-component models and $29.1\ {\rm rad\ m^{-2}}$ for three-component models (Figure~\ref{fig:psi0_RM_diff}). Although a small number of poorly constrained three-component fits produce extreme RM separations, the overall distributions are strongly concentrated toward low RM differences.

Taken together, the small intrinsic polarization angle and RM separations suggest that many of the fitted Faraday components are physically related, tracing emission regions embedded within a common or smoothly varying magneto-ionic environment rather than independent unresolved polarized structures \citep{OSullivan_2017}.

\subsection{Classification of Faraday complexity}

To investigate the dominant depolarization mechanisms in the LoTSS polarized source population, we classified the Faraday complexity of each source using the weighted depolarization parameters $\sigma_{\mathrm{RM,weighted}}$ and $\Delta \mathrm{RM}_{\mathrm{weighted}}$ together with their associated uncertainties. The parameter $\sigma_{\rm RM}$ traces turbulent fluctuations in Faraday depth and is typically associated with external Faraday dispersion \citep{burn66,Sokoloff98, Goodlet04, Mao14, Anderson15}, while $\Delta {\rm RM}$ traces coherent Faraday depth gradients or differential Faraday rotation, which may arise from internal Faraday rotating media or unresolved RM gradients across the source. For each source, the detection significance of the two parameters was computed as $|\sigma_{\rm RM,weighted}|/\sigma_{\rm RM,weighted,err}$ and $|\Delta {\rm RM}_{\rm weighted}|/\Delta {\rm RM}_{\rm weighted,err}$. A parameter was considered significantly detected if its significance exceeded the $3\sigma$ threshold. Sources for which neither $\sigma_{\rm RM}$ nor $\Delta {\rm RM}$ was significantly detected were classified as \textit{Faraday thin}. Sources with significant $\sigma_{\rm RM}$ but insignificant $\Delta {\rm RM}$ were classified as \textit{external screen dominated}, while sources with insignificant $\sigma_{\rm RM}$ but significant $\Delta {\rm RM}$ were classified as \textit{internal differential rotation}. Sources for which both parameters were significantly detected were classified as \textit{mixed internal/external} systems.

Applying this classification scheme to the sample of 1565 polarized sources shows that the LoTSS polarized population is strongly dominated by `external Faraday depolarization'. We find that 846 sources (54.06\%) are classified as external screen dominated, while 462 sources (29.52\%) are consistent with being Faraday thin across the observing band. Only 161 sources (10.29\%) are best described by internal differential rotation without significant external dispersion, and 96 sources (6.13\%) require both external and internal depolarization components simultaneously. Combining the mixed systems with the pure external and internal classes shows that signatures of external Faraday dispersion are present in 942 sources (60.2\%), whereas coherent Faraday depth gradients or differential Faraday rotation are present in only 257 sources (16.4\%).

These results indicate that the majority of polarized LoTSS sources are depolarized primarily by turbulent magneto-ionic material external to the synchrotron-emitting region. This behaviour is consistent with the classical Burn-law model of external Faraday dispersion \citep{burn66}, in which the fractional polarization decreases as $p(\lambda^2)=p_0\exp(-2\sigma_{\rm RM}^2\lambda^4)$. The strong dominance of significant $\sigma_{\rm RM}$ detections suggests that most polarized sources are embedded within or viewed through magnetized turbulent environments. In contrast, only a relatively small fraction of the sample requires significant $\Delta {\rm RM}$ without evidence for external Faraday dispersion. These systems may represent genuinely internally Faraday rotating sources, unresolved RM gradients across compact emitting regions, or complex jet structures, which are expected to occur more commonly in compact flat-spectrum AGN and blazar-like systems \citep{Sokoloff98, OSullivan_2012, Mao14, Pasetto18, Sokollf24}
. The large fraction of externally depolarized sources is also expected at LOFAR frequencies, since external Faraday dispersion scales strongly with wavelength ($\propto \lambda^4$), making low-frequency observations particularly sensitive to even modest RM fluctuations in foreground magnetized plasma \citep{burn66, Beck12, Anderson15, Heald15}.

\subsection{Relationship between Depolarization and Faraday Dispersion}
To investigate the relationship between spectral depolarization behaviour and Faraday complexity, we examined the dependence of the polarization spectral index, $\beta$, on the weighted Faraday dispersion parameter, $\sigma_{\mathrm{RM,wtd}}$, for sources containing different numbers of RM components. The parameter $\beta$ was obtained from power-law fits to the polarized intensity spectrum, $|P|=\sqrt{Q^2+U^2}$, as a function of wavelength squared using $P=A(\lambda^2)^{\beta}$, where \(\beta < 0\) indicates depolarization, while \(\beta > 0\) correlates with re-polarization \citep{OSullivan_2012, Farnes14, Anderson15, Pasetto18}. Increasingly negative values of $\beta$ therefore indicate stronger wavelength-dependent depolarization across the LOFAR HBA band.

\begin{figure}
    \centering
    \includegraphics[width=0.9\linewidth]{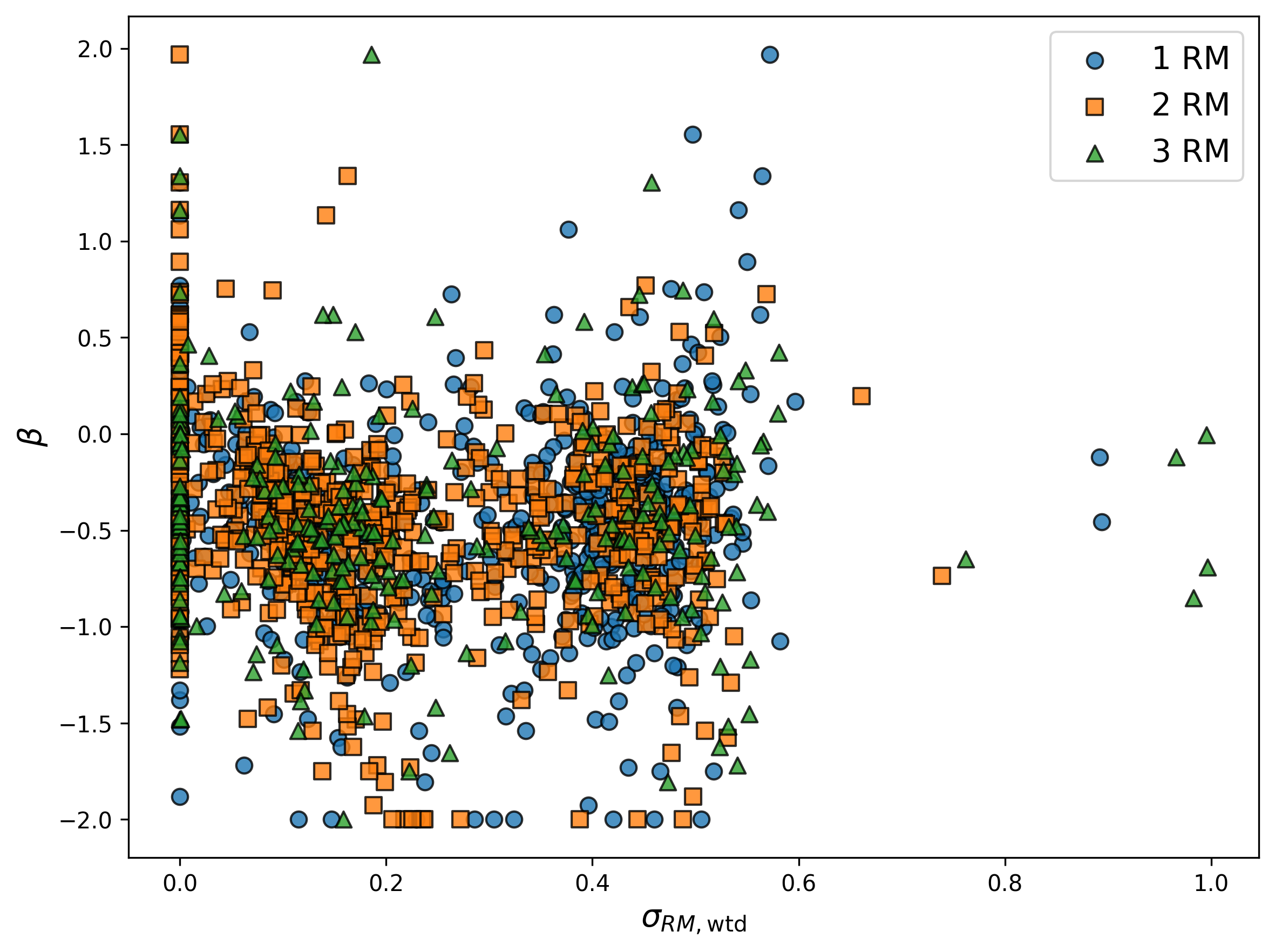}
    \caption{Polarization spectral index, $\beta$, versus weighted Faraday dispersion, $\sigma_{\mathrm{RM,wtd}}$, colour-coded by the number of RM components identified from the depolarization modelling. Negative $\beta$ values correspond to depolarization, while positive values indicate re-polarization. A weak but statistically significant anti-correlation is detected for the two-component sources, whereas the one- and three-component populations show no significant trend.}
    \label{fig:beta_sigmaRM}
\end{figure}

Figure~\ref{fig:beta_sigmaRM} presents \(\beta\) as a function of \(\sigma_{\mathrm{RM,wtd}}\), with the data points colour-coded according to the number of RM components identified through the depolarization modelling. For the single-component sources, no statistically significant correlation is found between $\beta$ and $\sigma_{\mathrm{RM,wtd}}$ ($r=-0.028$, $p=4.7\times10^{-1}$; $N=663$). The two-component sources show a weak but statistically significant anti-correlation ($r=-0.157$, $p=4.9\times10^{-5}$; $N=663$), indicating that increasing Faraday dispersion is associated with stronger depolarization behaviour in these systems. In contrast, the three-component sources show no significant correlation ($r=-0.013$, $p=8.3\times10^{-1}$; $N=278$). The large scatter in $\beta$ at low and intermediate $\sigma_{\mathrm{RM,wtd}}$ values suggests that the broad-band polarization behaviour cannot be explained solely by Faraday dispersion, and that multiple internal and external depolarization mechanisms likely contribute to the observed polarization spectra.

The observed relationship between the polarization spectral index, $\beta$, and the weighted Faraday dispersion parameter, $\sigma_{\mathrm{RM,wtd}}$, further supports the conclusion that external Faraday screens play a dominant role in shaping the polarization properties of the LoTSS source population. In the framework of external Faraday dispersion \citep{burn66}, increasing $\sigma_{\rm RM}$ corresponds to larger turbulent fluctuations in the foreground magneto-ionic medium, which produce progressively stronger wavelength-dependent depolarization. Consistent with this picture, the two-component sources exhibit a weak but statistically significant anti-correlation between $\beta$ and $\sigma_{\mathrm{RM,wtd}}$, indicating that sources with larger Faraday dispersion tend to display more negative $\beta$ values and therefore stronger depolarization across the LOFAR HBA band. This behaviour is qualitatively consistent with the dominance of external screen depolarization inferred from the Faraday complexity classification, where approximately 60\% of the sample shows evidence for significant external Faraday dispersion. The absence of a comparable trend in the one‑ and three‑component populations suggests that the connection between spectral depolarization and Faraday complexity is not universal, and that additional effects such as multiple unresolved RM components, internal Faraday rotation, beam depolarization, and complex source structure contribute to the observed scatter in polarization spectral behaviour \citep{Laing08,OSullivan_2012,Schnitzeler15,Heald15,Anderson15,Pasetto18}.

\subsection{Dependence of Faraday Dispersion and fractional mean Intrinsic Polarization on Source Size}
We investigated whether the rest-frame Faraday dispersion, $\sigma_{\rm RM,rest}$, and the intrinsic mean fractional polarization, $p_0$, depend on the projected linear size of the radio sources. Source sizes were characterized using the projected linear size measured in kpc, and correlations were evaluated separately for each Faraday class using both Pearson and Spearman rank statistics.  

No statistically significant correlation was detected between $\sigma_{\rm RM,rest}$ and source size in any population. Correlation coefficients for $\sigma_{\rm RM,rest}$ range from $r \sim -0.023$ (External screen dominated) to $r \sim 0.163$ (Mixed internal/external), with all corresponding p-values well above the 0.05 significance threshold. Similarly, the intrinsic mean fractional polarization, $<p_0>$, shows no significant correlation with source size, with correlation coefficients ranging from $r \sim -0.065$ (Internal differential rotation) to $r \sim 0.049$ (Mixed internal/external), and all p-values indicate no statistically significant associations. 

The absence of a strong size dependence suggests that the magneto-ionic environments responsible for Faraday depolarization and intrinsic polarization are not primarily regulated by the evolutionary expansion of the radio source \citep{Strom73, Pedelty89, IshwaraChandra98}. Classical depolarization models predict that larger and more evolved radio galaxies should exhibit reduced Faraday dispersion and potentially higher fractional polarization as their lobes propagate beyond the dense interstellar medium into lower-density circumgalactic environments. However, our results imply that environmental scatter and large-scale magnetized structures dominate over any simple size-driven evolutionary trend \citep{Goodlet04, Lamee16}.  

The lack of a significant $<p_0>$ – linear size correlation further indicates that the intrinsic ordering of magnetic fields, as traced by fractional polarization, does not strongly scale with the physical size of the source. Combined with the strong redshift dependence observed in partial correlation analyses (Section \ref{sec:redepol}), this suggests that cosmic evolution of the surrounding magnetized medium is a more significant driver of both $\sigma_{\rm RM,rest}$ and $p_0$ than projected source size.  

We note that these results may partly reflect observational limitations associated with the spatial resolution of polarization measurements. Many sources are resolved by only a few synthesized beams, such that measured depolarization and polarization properties represent beam-averaged quantities rather than spatially resolved Faraday structures along the radio lobes. Classical size–depolarization and size–polarization trends are expected to arise from environmental gradients along the lobes; unresolved or marginally resolved sources may suppress intrinsic correlations by averaging internal Faraday variations. Consequently, the lack of a strong correlation should not necessarily be interpreted as evidence against environmental evolution with source growth, but rather as an indication that \textit{higher-resolution, spatially resolved polarization measurements} are required to fully recover these trends.

\section{Redshift and Luminosity Evolution of Faraday Depolarization}
\label{sec:redepol}
To explore the evolution of Faraday depolarization properties as a function of cosmic epoch and radio luminosity, we examined the dependence of the rest-frame Faraday dispersion, $\sigma_{\rm RM,rest}$, on redshift and radio luminosity for each Faraday class. The observed RM dispersion was transformed to the source rest frame following the expected redshift scaling of Faraday rotation, $\sigma_{\rm RM,rest} = \sigma_{\rm RM,obs}(1+z)^2$, where $\sigma_{\rm RM,obs}$ is the observed-frame RM dispersion and $z$ is the source redshift.

Correlations were quantified using both Pearson and Spearman rank statistics. We find statistically significant positive correlations between $\sigma_{\rm RM,rest}$ and redshift for the \textit{External screen dominated}, \textit{Faraday thin}, and \textit{Mixed internal/external} populations. The strongest trend is observed for the \textit{Mixed internal/external} class ($r_{\rm S}=0.693$, $p=1.55\times10^{-11}$), followed by the \textit{External screen dominated} sources ($r_{\rm S}=0.626$, $p=1.50\times10^{-76}$). The \textit{Faraday thin} population also shows a weaker but statistically significant correlation ($r_{\rm S}=0.315$, $p=6.35\times10^{-6}$). Similar positive trends are observed between $\sigma_{\rm RM,rest}$ and radio luminosity for these same populations. In contrast, the \textit{Internal differential rotation} class does not exhibit statistically significant Spearman correlations with either redshift or luminosity, despite moderately large Pearson coefficients, suggesting that the apparent trends may be influenced by a small number of extreme sources rather than representing a robust monotonic evolution \citep{Laing08,OSullivan_2012,Mao14}. 

\begin{figure}
    \centering
    \includegraphics[width=0.95\linewidth]{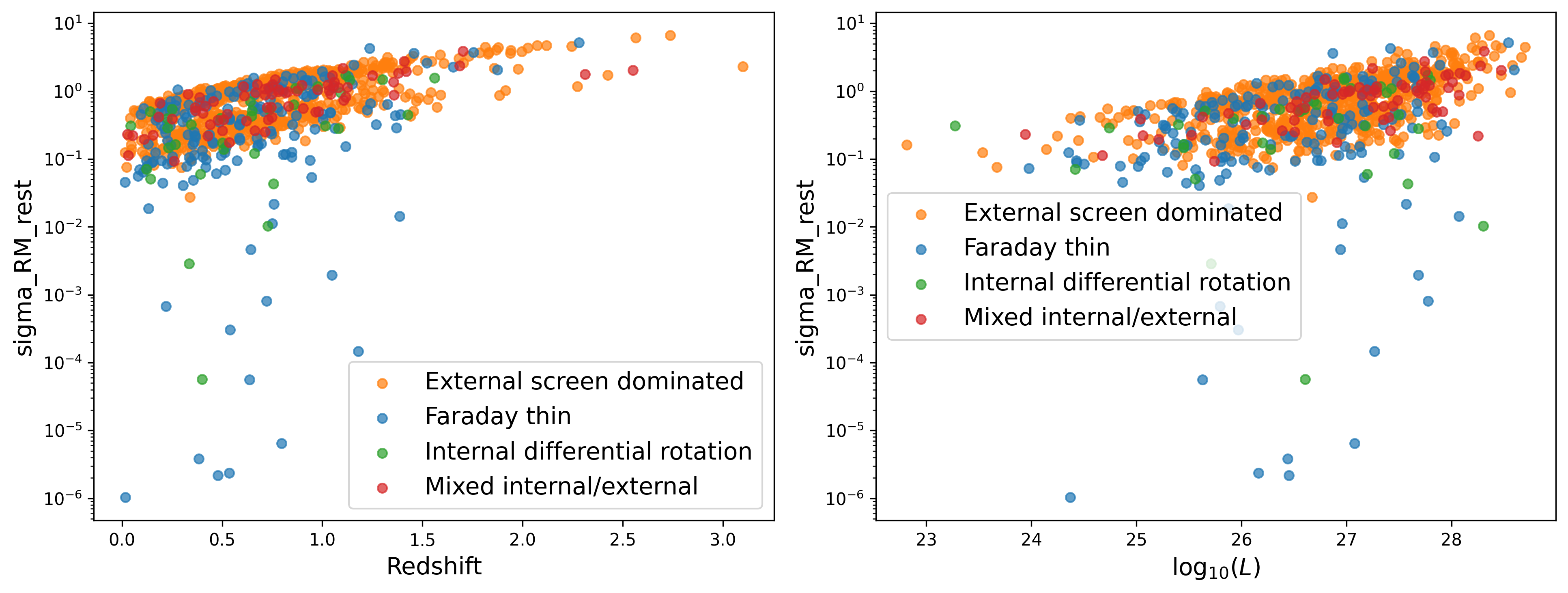}
    \caption{Evolution of the rest-frame Faraday dispersion, $\sigma_{\rm RM,rest}$, with redshift (left panel) and radio luminosity (right panel) for the different Faraday depolarization classes. The observed RM dispersion has been corrected to the source rest frame using $\sigma_{\rm RM,rest}=\sigma_{\rm RM,obs}(1+z)^2$. The \textit{External screen dominated}, \textit{Faraday thin}, and \textit{Mixed internal/external} populations exhibit significant positive correlations with both redshift and luminosity, indicating increasingly turbulent and magnetized environments at higher redshifts and radio powers. The y-axis is shown in logarithmic scale.}
    \label{fig:sigmaRM_z_L}
\end{figure}

The observed increase of $\sigma_{\rm RM,rest}$ with redshift and luminosity indicates that sources dominated by external Faraday-active media are embedded within increasingly turbulent, dense, or magnetized environments at earlier cosmic epochs and for higher intrinsic radio powers \citep{burn66,Beck12,Heald15, Rappaz24}. This behaviour is consistent with depolarization arising predominantly from magneto-ionic material external to the synchrotron-emitting region, such as turbulent circumgalactic or intracluster media \citep{Laing08,Anderson15,Pasetto18}. For comparison, the weighted RM difference parameter, $\Delta {\rm RM_{wtd}}$, does not show statistically significant correlations with either redshift or luminosity for any Faraday class, implying that localized RM asymmetries or gradients remain comparatively insensitive to these large-scale evolutionary effects \citep{OSullivan_2012,Mao14}. Consequently, only the $\sigma_{\rm RM,rest}$ evolution trends are included in the Figure~\ref{fig:sigmaRM_z_L}.

\subsection{Partial Correlation Analysis: Disentangling Redshift and Luminosity Effects}

To disentangle the intrinsic dependence of the rest-frame Faraday dispersion, $\sigma_{\rm RM,rest}$, on redshift and radio luminosity, we performed a partial correlation analysis to account for the strong covariance between these quantities in flux-limited radio surveys. In such samples, redshift and luminosity are intrinsically coupled due to selection effects, where low-luminosity sources are preferentially detected only at low redshift, while high-redshift populations are dominated by intrinsically luminous AGN. Consequently, apparent bivariate correlations may not reflect true physical dependencies.

To mitigate this degeneracy, we computed partial rank correlations by removing the contribution of the control variable through linear regression residual analysis. Specifically, for a given pair of variables $X$ and $Y$ with control variable $Z$, we model the linear dependence as
\[
X = a_X + b_X Z + \epsilon_X, \quad
Y = a_Y + b_Y Z + \epsilon_Y,
\]
where $\epsilon_X$ and $\epsilon_Y$ represent the residuals after removing the dependence on $Z$. The partial correlation coefficient is then estimated as the Spearman rank correlation between the residuals,
\[
\rho_{XY \cdot Z}
=
\mathrm{Spearman}(\epsilon_X,\epsilon_Y).
\]

In this work, we applied this framework in two configurations: (i) assessing the dependence of $\sigma_{\rm RM,rest}$ on redshift while controlling for radio luminosity, $\rho_{\sigma_{\rm RM,rest},\, z \, \cdot \, \log L}$, and (ii) assessing the dependence of $\sigma_{\rm RM,rest}$ on luminosity while controlling for redshift, $\rho_{\sigma_{\rm RM,rest},\, \log L \, \cdot \, z}$.

We find that the partial correlation between $\sigma_{\rm RM,rest}$ and redshift remains statistically significant across all Faraday classes after controlling for luminosity, with the strongest signal observed in the \textit{Mixed internal/external} population ($\rho \approx 0.59$, $p \sim 6 \times 10^{-8}$) and the \textit{External screen dominated} class ($\rho \approx 0.45$, $p \sim 10^{-35}$). The \textit{Faraday thin} and \textit{Internal differential rotation} populations also retain weaker but significant residual redshift trends. In contrast, when controlling for redshift, the partial correlation between $\sigma_{\rm RM,rest}$ and radio luminosity becomes weak or statistically insignificant in most classes, remaining only marginally significant for the \textit{External screen dominated} sources and consistent with no correlation in the other populations.

These results demonstrate that the observed evolution of $\sigma_{\rm RM,rest}$ is primarily driven by redshift, indicating a genuine increase in Faraday-active magneto-ionic environments toward earlier cosmic epochs \citep{Beck12,Heald15, Rappaz24}. The weaker and less consistent luminosity dependence suggests that more powerful radio AGN may reside in more magnetized or turbulent environments, but this effect is secondary to cosmic evolution \citep{Anderson15,Pasetto18}. Overall, the persistence of the redshift dependence after removing luminosity covariance provides strong evidence for intrinsic cosmological evolution in the magnetized plasma surrounding radio AGN, rather than a selection-induced artefact of flux-limited sampling \citep{Browne_Marcha_93, Wall97}.

\section{Summary and Conclusions}
\label{sec:discuss}
We have presented a comprehensive broadband polarimetric analysis of 1565 LoTSS-DR2 sources, modelling their Stokes \(Q\) and \(U\) spectra across the LOFAR HBA band using multi-component Faraday depolarization models. The goal of this work was to quantify the Faraday complexity of the polarized source population and to constrain the dominant physical mechanisms responsible for depolarization in low-frequency radio sources.

\subsection{Summary of main results}

The best-fitting models provide good descriptions of the observed polarization behaviour, with reduced \(\chi_r^2\) values typically close to unity (median \(\chi_r^2 \approx 1.51\)). A substantial fraction of the sample requires multiple Faraday components, with 43.2\% of sources best described by two or three components, indicating widespread complexity in the magneto-ionic medium. Only 16.2\% of sources are consistent with being Faraday thin, showing no measurable Faraday dispersion or differential Faraday rotation.

External Faraday dispersion is the dominant depolarization mechanism in the sample: 54.1\% of sources are classified as external screen dominated, and 60.2\% show statistically significant evidence for external Faraday dispersion. In contrast, 16.4\% exhibit signatures of differential Faraday rotation without external dispersion, and 6.1\% require a combination of both effects. These results indicate that turbulent foreground or surrounding magnetoionic media predominantly shape the observed polarization at LOFAR frequencies.

The distributions of Faraday parameters are generally low to moderate in amplitude, with median polarization-weighted values of \(\sigma_{\mathrm{RM,wtd}} \sim 0.3~\mathrm{rad\,m^{-2}}\) and \(\Delta \mathrm{RM}_{\mathrm{wtd}} \sim 0.5~\mathrm{rad\,m^{-2}}\). The polarization spectral index \(\beta\) spans a broad range from approximately \(-2\) to \(+2\), with a median of \(\beta_{\rm med} \approx -0.44\), indicating that most sources undergo net depolarization across the LOFAR HBA band.

\subsection{Physical interpretation of Faraday structure}

Multi-component Faraday modelling suggests that RM components are frequently physically related. Intrinsic polarization angle differences typically lie \(\lesssim 30^\circ\), and pairwise RM separations peak at \(\sim 20~\mathrm{rad\,m^{-2}}\), indicating that many fitted components trace emission embedded within a common or smoothly varying magneto-ionic environment rather than independent, uncorrelated polarized structures. This supports a picture in which multiple Faraday components arise from emission regions embedded in a shared foreground or surrounding magnetized medium.

The predominance of external Faraday dispersion is consistent with classical turbulent screen models, where depolarization scales as \(\lambda^4\). This implies that a large fraction of the polarized emission is observed through or embedded in magnetized plasma such as intragroup or intracluster gas, or extended halos associated with radio galaxies and AGN \citep{Sokoloff98, Laing08, Beck13, Vanweeren19}. The modest median values of \(\sigma_{\mathrm{RM,wtd}}\) and \(\Delta \mathrm{RM}_{\mathrm{wtd}}\) further suggest that, on average, the Faraday structure is not highly turbulent but still sufficient to drive significant broadband depolarization at LOFAR frequencies.

\subsection{Spectral depolarization and Faraday complexity}

The relationship between the polarization spectral index \(\beta\) and weighted Faraday dispersion \(\sigma_{\mathrm{RM,wtd}}\) is generally weak, but a statistically significant anti-correlation is detected for two-component sources. In this subpopulation, larger Faraday dispersion corresponds to more negative \(\beta\), reflecting stronger wavelength-dependent depolarization. This trend is qualitatively consistent with external Faraday dispersion as the primary driver of spectral depolarization.

However, the absence of a consistent correlation in the one- and three-component populations, together with substantial scatter at low and intermediate \(\sigma_{\mathrm{RM,wtd}}\), indicates that spectral depolarization cannot be explained by Faraday dispersion alone. Additional mechanisms—including internal Faraday rotation, beam depolarization, unresolved RM substructure, and complex source geometry—likely contribute to the observed diversity of polarization spectra. This highlights the importance of full spectro‑polarimetric modelling, as broadband polarized intensity spectra alone are insufficient to fully characterise Faraday complexity.

\subsection{Evolution with redshift and radio power}

We find significant positive correlations between rest-frame Faraday dispersion \(\sigma_{\rm RM,rest}\) and both redshift and radio luminosity for sources dominated by external Faraday effects. These correlations remain robust after accounting for the covariance between redshift and luminosity via partial correlation analysis, indicating that redshift is the primary driver of the observed evolution. In contrast, the RM gradient term \(\Delta \mathrm{RM}\) shows no significant dependence on either redshift or luminosity, suggesting that coherent Faraday depth gradients are less sensitive to large-scale environmental evolution.

The increase of $\sigma_{\rm RM,rest}$ with redshift implies that radio sources at earlier cosmic epochs tend to reside in more turbulent, dense, or strongly magnetized environments. This is consistent with evolving circumgalactic and intra-cluster media and supports a scenario in which external magneto-ionic environments play an increasingly important role in shaping the observed polarization properties of radio AGN at high redshift. Radio luminosity appears to have a weaker, secondary influence, with the strongest environmental trends observed in the most luminous systems. In contrast, no statistically significant correlation was detected between $\sigma_{\rm RM,rest}$ and projected radio source size for any Faraday class.

\subsection{Conclusions}

\begin{enumerate}
  \item External Faraday dispersion emerges as the dominant depolarization mechanism within the LoTSS polarized source population, with turbulent foreground or surrounding magnetoionic media shaping most polarization spectra at LOFAR frequencies.

  \item Most sources exhibit multi-component Faraday structure, with RM components that are typically physically related and embedded in a common or smoothly varying magneto-ionic environment rather than arising from independent polarized regions.

  \item Polarization spectral behaviour alone is insufficient to fully characterise Faraday complexity, full spectro‑polarimetric modelling of \(Q\) and \(U\) across the band is required to disentangle distinct depolarization mechanisms.

  \item Rest-frame Faraday dispersion \(\sigma_{\rm RM,rest}\) increases significantly with redshift, indicating that radio sources at earlier cosmic epochs are embedded in stronger or more turbulent magnetoionic environments, consistent with evolving circumgalactic and intra-cluster media.

  \item Internal Faraday rotation is comparatively rare, with only a small fraction of sources requiring significant differential rotation without strong external Faraday dispersion, supporting the dominance of external turbulent screens in the overall source population.

  \item The lack of a significant $\sigma_{\rm RM,rest}$--size correlation indicates that projected source size is not the primary driver of Faraday depolarization in the present sample, with redshift evolution appearing to dominate over simple source-growth effects.
    
\end{enumerate}

Overall, this work demonstrates that low‑frequency polarimetry provides powerful constraints on the magnetoionic environments of radio sources across cosmic time. The strong prevalence of external Faraday dispersion and its evolution with redshift highlight the importance of environment in shaping polarization properties, and set the stage for future high‑resolution, multi‑frequency studies aimed at disentangling internal and external Faraday effects in individual systems.

\section*{Acknowledgements}
 RS acknowledges Banwarilal Bhalotia College, under Kazi Nazrul University, where this work was carried out as part of the M.Sc.\ Physics programme. The authors acknowledge ChatGPT for assistance with copy editing and improving manuscript clarity. AG acknowledges support from the SERB-SURE grant (SUR/2022/000595), DST, Government of India, and thanks IUCAA, Pune for support through the Associateship Programme and access to computational facilities.

\section*{Data Availability}
The RM Grid catalogue, together with the corresponding \(Q\) and \(U\) frequency spectra for each catalogue entry, is publicly available at \url{https://lofar-mksp.org/data/}.
The full set of LoTSS-DR3 images, best-fit model parameters, plots, and tables for all 1565 sources are publicly available at
\url{https://drive.google.com/drive/folders/1sUXxFDrMvyVSivq9QoaYfvgftTsnJ2Ko?usp=sharing}.

%%%%%%%%%%%%%%%%%%%% REFERENCES %%%%%%%%%%%%%%%%%%
\bibliographystyle{mnras}
\bibliography{main} 

%%%%%%%%%%%%%%%%%%%%%%%%%%%%%%%%%%%%%%%%%%%%%%%%%%

%%%%%%%%%%%%%%%%% APPENDICES %%%%%%%%%%%%%%%%%%%%%

% Don't change these lines
\bsp	% typesetting comment
\label{lastpage}
\end{document}